\documentclass[preprint2]{aastex}

\usepackage{color}

\slugcomment{To appear in ApJL}

\shorttitle{Methods to Determine if Exo-moon Orbit is Prograde or Retrograde}
\shortauthors{Lewis \& Fujii}

\begin{document}

\title{Next Generation of Telescopes or Dynamics Required to Determine if Exo-Moons have Prograde or Retrograde Orbits}

\author{Karen M. Lewis and Yuka Fujii}
    
\affil{Earth-Life Science Institute (WPI-ELSI), Tokyo Institute of Technology, Ookayama, Meguro district, Tokyo 152-8551, Japan}

\begin{abstract}
We survey the methods proposed in the literature for detecting moons of extrasolar planets in terms of their ability to distinguish between prograde and retrograde moon orbits, an important tracer of moon formation channel.  We find that most moon detection methods, in particular, sensitive methods for detecting moons of transiting planets, cannot observationally distinguishing prograde and retrograde moon orbits.  The prograde and retrograde cases can only be distinguished where dynamical evolution of the orbit due to e.g. three body effects is detectable, where one of the two cases is dynamically unstable or where new observational facilities which can implement a technique capable of differentiating the two cases, come on line.  In particular, directly imaged planets are promising targets as repeated spectral and photometric measurements, required to determine moon orbit direction, could also be conducted with the primary interest of characterising the planet itself.
\end{abstract}


\section{Introduction}

As a result of projects such as SEEDS \citep{Tamura2009}, aimed at finding new directly imaged planets, the launch of  satellites such as Kepler \citep{Borucki2008} and the consequent wealth of candidate transiting planet systems, much scientific effort has been directed towards follow-up.  For the case of directly imaged planets this could involve constraining orbits through astrometry and dynamics \citep[e.g][]{Kalas2008} and determining atmospheric composition \citep[e.g.][]{Konopacky2013}, while for transiting planets possible follow-up includes using dynamical \citep[e.g.][]{Lissauer2011}  perturbations to confirm planetary masses, and searching for moons of transiting planets \citep{Kipping2012}.

The idea of detecting moons of extrasolar planets is not new and many detection methods have been proposed.  For the case of transiting planets, moons can be detected through perturbation of the transit lightcurve through the extra dip due to the moon \citep[e.g.][]{Sartoretti1999}, transit timing \citep[e.g.][]{Sartoretti1999,Szabo2006,Kipping2009a,Kipping2009b} and duration variations \citep{Kipping2009a,Kipping2009b}, changes in mean \citep{Heller2014} and scatter \citep{Simon2012} in stacked lightcurves and perturbation in the Rossiter-McLaughlin signal due to the planetary transit \citep{Simon2010}.  For the case of directly imaged planets, moon detection has been proposed by monitoring the image of the planet for variations in infra-red \citep{Moskovitz2009,Peters2013} and visual \citep{Cabrera2007} luminosity, spectral features that could only be due to a moon \citep{Williams2004} and directly, for motion about the planet-moon barycenter using position and velocity information \citep{Cabrera2007}.  In addition, it has also been proposed that moons of microlensed planets \citep{Han2002,Han2008}, radial velocity planets \citep[e.g][]{Podsiadlowski2010} and pulsar planets \citep{Lewis2008} can be detected through perturbation of the planetary signal.

In addition, extrapolating from the Solar System, moons of exoplanets could be numerous.  This idea is supported by the recent possible detection of a moon of a microlensed planet \citep{Bennett2013}.  While this detection cannot be confirmed, moon formation may be common, as a number of directly detected young planets have spectra  consistent with a circumplanetary disk of material \citep[e.g.][]{Kalas2008}. 

\section{What Exo-Moon Properties Tell Us}
\label{Moon_Formation}

The physical and orbital properties of satellite systems of extrasolar planets allow tests of planet and moon formation models.  In particular, recent work on the formation/origin of impact generated moons of terrestrial planets, regular satellites of gas giants and captured moons yield a set of physically motivated limits on moon mass, semi-major axis and inclination, which will be summarised in turn.

For terrestrial planets with mass less than 2.5 Earth masses, a giant impact is sufficiently energetic to place up to 4\% of the planet's mass into orbit \citep[e.g.][]{Canup2001}, but not energetic enough to disperse this disk \citep{Wada2006}.  From this disk, one or more protomoons form, but due to gravitational perturbation from planetary tides \citep{Atobe2007} or between protomoons \citep{Canup1999}, only systems with single moons on coplanar orbits, or close, inclined orbits survive e.g. planet-moon or moon-moon collisions.

Alternatively, gas giant regular satellites are thought to form in a prograde (i.e. in the same direction as the planet orbit) circumplanetary disk \citep[e.g.][and references therein]{Tanikawa2014}, fed from the circumstellar disk.  The prograde motion of such a circumplanetary disk is driven by inward gas motion in the leading horseshoe region and outward gas motion in the trailing horseshoe region, motions resulting from these regions connecting the slower moving outer part and the faster moving inner part of the circumstellar disk. Such a system yields satellite orbits aligned with the planetary spin axis, which can be subsequently modified by e.g. secular spin-orbit interactions \citep[e.g.][]{Ward2004}.  Models of such disks have succeeded in reproducing close-in, multiple moon systems like those found in nature, in particular, in terms of satellite location \citep{Mosqueira2003a,Sasaki2010} and total moon mass \citep{Canup2006,Sasaki2010}.

Finally, planets can capture large moons.  The orbit direction of a captured body depends sensitively on the approach trajectory, with both retrograde (i.e. in the opposite direction to the planet orbit) and prograde orbits possible \citep[e.g.][fig. 4]{Tanikawa2014}.  Additionally, for distant moons, moons on retrograde orbits survive longer than those on prograde orbits due to their increased relative stability \citep[e.g][]{Hamilton1991}.  Consequently, a mix of prograde and retrograde orbits are expected for captured moons \citep{Agnor2006,Ochiai2014}, with physical properties depending on the capture pathway, e.g. Triton-like moons for three-body capture \citep{Agnor2006} and gas giant binaries for tidal capture \citep{Ochiai2014}.

Consequently, physical properties of a moon e.g. mass, can be used to test a given formation model while orbital properties, in particular, if the moon's orbit  is prograde or retrograde, can differentiate between different formation models e.g. regular satellite or captured satellite.

\begin{figure*}[t]
\includegraphics[scale=.55]{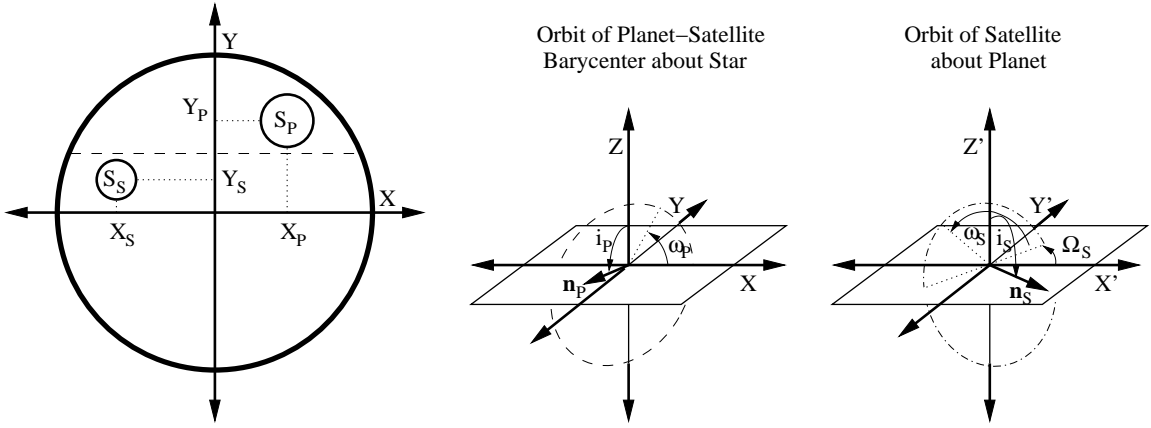}
\caption{Schematic diagram of coordinate system.  The left panel shows the position of the planet (large circle) and satellite (small circle) on the face of the star during transit.  The right two panels show the orientation of the barycenter (dashed) and satellite (dot-dashed) orbits in terms of the Euler angles, where $\mathbf{n}$ is a unit vector normal to the orbital plane.  Following \citet{Kipping2011}, we set $\Omega_P = 0$, and define the reference plane and direction for the satellite orbit to be the plane and direction of pericenter of the barycenter orbit.}
\label{coordinate_system}
\end{figure*}

\section{Which Moon Detection Techniques Distinguish Between Prograde and Retrograde Moon Orbits?}

\subsection{Moons of Transiting Planets}

To derive the shape of the transit lightcurve for a planet-moon pair as a function of time, we follow  \citet{Gimenez2006} and write
\begin{eqnarray}
&L = L_0 - \int_{S_{P}(t)} I(\mu_P)dA - \int_{S_{S}(t)} I(\mu_S)dA \nonumber\\
& + \int_{S_{SP}(t)} I(\mu_S)dA, \label{prelim_coord_xpdef}
\end{eqnarray}
where $L$ and $L_0$ are the measured and out-of-transit luminosities, $I$ is the intensity on the face of the star as a function of position, $S_P$ and $S_S$ are the regions of the star occulted by the planet and satellite respectively (see figure \ref{coordinate_system}), and $S_{PS}$ is the overlap between $S_P$ and $S_S$, where applicable.

To determine the dependance of $S_S(t)$ and $S_P(t)$, and thus $L(t)$ on time, expressions for the positions of the planet and moon on the plane of the sky are needed.  Following \citet{Kipping2011}, the position of the planet and moon on the face of the star are given by
\begin{eqnarray}
& X_P = r_P \cos(f_P + \omega_P) \nonumber\\
&- r_{PB} \left[\cos(\omega_P + \Omega_S) \cos(f_S + \omega_S) \right. \nonumber\\
& \left. - \cos i_S \sin (\omega_P + \Omega_S)  \sin(f_S + \omega_S)\right], \label{prelim_coord_xpdef}
\end{eqnarray}
\begin{eqnarray}
&X_S = r_P \cos(f_P + \omega_P) \nonumber\\
&+ r_{SB} \left[\cos(\omega_P + \Omega_S) \cos(f_S + \omega_S) \right. \nonumber\\
& \left. - \cos i_S \sin (\omega_P + \Omega_S)  \sin(f_S + \omega_S)\right], \label{prelim_coord_xmdef}
\end{eqnarray}
\begin{eqnarray}
&Y_P = r_P \cos i_P \sin(f_P + \omega_P) \nonumber \\
&- r_{PB} [[-\sin i_S \sin i_P \nonumber \\
&   + \cos i_S \cos i_P \cos (\omega_P + \Omega_S)] \sin(f_S + \omega_S) \nonumber \\
&  + \cos i_P \sin(\omega_P + \Omega_S) \cos(f_S + \omega_S) ], \label{prelim_coord_ypdef}
\end{eqnarray}
\begin{eqnarray}
&Y_S = r_P \cos i_P \sin(f_P + \omega_P) \nonumber \\
&+ r_{SB} [[-\sin i_S \sin i_P \nonumber \\
&   + \cos i_S \cos i_P \cos (\omega_P + \Omega_S)] \sin(f_S + \omega_S) \nonumber \\
&  + \cos i_P \sin(\omega_P + \Omega_S) \cos(f_S + \omega_S) ], \label{prelim_coord_ymdef}
\end{eqnarray}
where $X_P$ and $Y_P$, and $X_S$ and $Y_S$ are the $X$ and $Y$ coordinates of the planet and satellite respectively on the face of the star (see figure~\ref{coordinate_system}), $f_P(t)$ and $f_S(t)$ are the true anomalies of the orbit of the planet-satellite barycenter and the satellite, and $r_P(t)$, $r_{PB}(t)$ and $r_{SB}(t)$ are the distances between the planet-satellite barycenter and the system barycenter, planet and satellite respectively.  In addition, $\omega_P$, $\omega_S$ , and $i_P$ and $i_S$ are the arguments of perihelion, and inclinations of both orbits, where we note that the reference plane of the satellite's orbit is the planet's orbit.  Also, $\Omega_S$ is the longitude of the ascending node of the satellite's orbit, where, following \citet{Kipping2011},  we set $\Omega_P = 0$ as this does not affect the analysis.

\begin{figure}[tb]
\includegraphics[scale=.14]{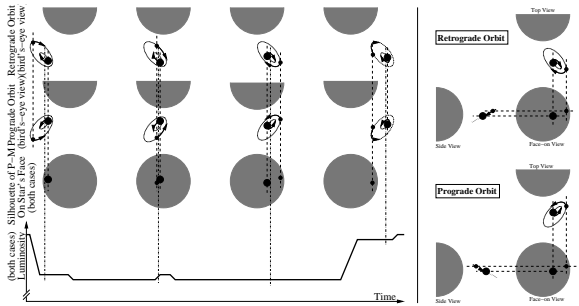}
\caption{Diagram of the position of the planet and moon (black dots) on the face of the star (grey) for four epochs during transit along with the retrograde and prograde orbits capable of producing the observed transit lightcurve (below).  The center of mass of the planet-moon system is shown with a dot-dash line while the orbits of the planet and moon are dashed.  
In the right panel the prograde and retrograde orbits are shown in orthographic projection.}
\label{ReflectionFig}
\end{figure}

To see that transit lightcurves cannot be used to distinguish between prograde and retrograde moon orbits, consider the case of a prograde moon orbit (see second row of figure~\ref{ReflectionFig}).  By reflecting this orbit across the plane of the sky ($X$-$Y$ plane) it can be transformed into a different retrograde orbit (see first row of figure~\ref{ReflectionFig}).  Also, as this is a reflection across the $X$-$Y$ plane, the $X$ and $Y$ coordinates of the planet and satellite will be the same for both the prograde and retrograde cases.  As the time dependance of $L(t)$ is entirely determined by the time dependance of $S_P$ and $S_S$, which is in turn determined by the time dependances of $X_P$, $Y_P$, $X_S$ and $Y_S$ (which will be the same for both cases), both systems will produce \emph{identical} lightcurves.  

In this coordinate system there is no general analytic transformation between the orbital elements of the prograde and retrograde orbits capable of producing the same light curve, except in special cases.  For example, when $\omega_P + \Omega_S = 0$, the orbital elements of the retrograde orbit can be found from those of the prograde orbit and visa versa by replacing $i_S$ with $2\pi-i_P - (i_S+i_P)$.  As $ i_P \approx \pi/2$ in order for the planet to transit, it can be seen that if the original satellite orbit inclination was between $0$ and $\pi/2$ (prograde), the new inclination will, except for very face on orbits, be between $\pi/2$ and $\pi$ (retrograde).  To test that this result was general, a perl program\footnote{Program available on request.} was written to calculate the orbital elements of a reflected orbit from the orbital elements of a known orbit.  

\citet{Kipping2009b, Kipping2011} suggested that transit duration variation could be used to differentiate between prograde and retrograde moon orbits.  We agree with the result that the light curve resulting from a given satellite orbit, and the same orbit, but with the direction reversed will generally not produce the same transit light curves, however, the transformation we present is not merely a reversal of the direction of the orbit, as was the case investigated in these works.  In particular, we suggest that the light curve produced by a prograde orbit can be matched by a another retrograde orbit with the same shape, but different orientation.

Consequently, moon detection methods which rely on the time dependance of $S_P$ and $S_S$, including, lightcurve distortion \citep{Sartoretti1999,Szabo2006,Simon2012,Heller2014}, variations in transit mid-time \citep{Sartoretti1999} or duration \citep{Kipping2009a} or Rossiter-McLaughlin perturbation cannot distinguish between the prograde and retrograde cases.    Finally, mutual events, that is, bumps in the lightcurve due to the moon passing in front of or behind the planet during transit, while detectable, are only useful for determining orbit direction in extreme cases.  In particular, for the case where the radiation from the planet and moon is detectable \emph{and} where the moon has a detectably different brightness from the planet, such that the case of a moon passing in front of and behind a planet can be differentiated, mutual events can be used to differentiate between the retrograde and prograde cases.  This would be very technically challengeing as TPF-C was capable of detecting such events for the case of directly imaged planets in the presence of speckle noise \citep{Cabrera2007} as opposed to much higher amplitude stellar photon noise in the transiting planet case.  

Two options exist that may resolve the degeneracy in the transit light curve.  First, if the system is observed for a sufficient time, dynamical perturbation to the satellite orbit, e.g. from the host star may allow the prograde and retrograde cases to be distinguished.  For example, \citet{Carter2011} used dynamical perturbations in binary eccentricity and inclination of amplitude $\sim$$0.005$ and $8^{\circ}$ with period 650 and 950 days, to determine a unique orbit model for a stellar binary which orbited and transited a third star.  The amplitude and period of these dynamically induced perturbations depends non-trivially on the star, planet and moon masses, and the moon and planet-moon barycenter orbits \citep[e.g.][eq. 48 \& 49]{Mardling2013}.  As information on e.g. the orientation and eccentricity of the barycenter orbit is known from radial velocity measurements, these perturbations may allow the prograde and retrograde cases to be differentiated, however, measuring these small changes may be extremely challenging.  Second, the planet's line-of-sight velocity, measured using atmospheric absorption lines \citep[e.g.][]{Snellen2010}, can be combined with light curve information, as the line-of-sight velocity of the planet is different for prograde and retrograde orbits (see figure~\ref{ReflectionFig}).  The observations made by \citet{Snellen2010} were of a very bright star using a 8.2m telescope.  The calculated error in planet velocity (10 kms$^{-1}$) is an order of magnitude larger than the possible signal due to a realistic moon (see table~\ref{SigTable}).  Thus, this approach is only feasible if the next generation of telescopes are built.

\subsection{Moons of Microlensed Planets}

Moons of microlensed planets may be detectable for moons larger than an Earth mass with projected planet-moon distances  larger than a planetary Einstein radius \citep{Han2008}.  However, microlensing is only sensitive to the projected separation between the planet and the moon  \citep[e.g.][]{Han2008}.  Consequently, as with the transit technique, this technique cannot  differentiate between prograde and retrograde moon orbits.

\subsection{Moons of Radial Velocity and Pulsar Planets}

Moons of radial velocity planets and pulsar planets can be detected through the additional perturbation to the radial velocity or time-of-arrival signal of the host star, resulting from planet-moon binarity \citep{Podsiadlowski2010,Lewis2008}.   These techniques can distinguish between prograde and retrograde moon orbits, however, they  are not very useful as they are only capable of detecting wide gas giant binaries and the perturbation due to a binary pair may be mistaken for the signal from another planet \citep{Schneider2006}.  For the case of pulsar planets, progress can be made by finding new millisecond pulsars, or by reducing the timing noise using new facilities such as the square kilometer array (SKA).  For a PSR B1620-26b analog planet, assuming $\sim$100ns timing noise for the SKA \citep{Liu2011}, detection of distant ($a_S <$ 0.5AU) Earth-mass moons may be possible \citep[][eq. 14]{Lewis2008}.  For the case of radial velocity planets, Jupiter mass planet pairs cause a radial velocity perturbation on the order of 0.3cm$^{-1}$ \citep{Podsiadlowski2010}, well below current sensitivity and would require new instruments and technologies, such as those proposed by \citet{Lovis2006} for an OWL\footnote{The OverWhelmingly Large telescope (100m diameter).}-like telescope.

\subsection{Moon of Directly Imaged Planets}

For directly imaged planets, six moon detection techniques have been proposed.  However, only three can distinguish between prograde and retrograde orbits.  In particular, method~1, spectrally detecting terrestrial moons through the methane window of a host giant planet's spectrum \citep{Williams2004} only gives information on the presence or absence of an Earth-like moon.  Similarly, method 2, detecting a moon using IR thermal variation over a planetary year (due to a moon's low thermal inertia) \citep{Moskovitz2009}, and method 3, detecting radiation from a thermally heated moon \citep{Peters2013}, again only indicate a moon's presence.  Alternatively, method 4, proposed by \citet{Cabrera2007} involving searching for mutual events in the infra red or optical can distinguish between a planet passing in front of a moon and a moon passing in front of a planet, and it has been suggested that such events could allow the detection of lunar-sized moons of Earth analogs by missions similar to TPF-C.  In addition, \citet{Cabrera2007} also proposed that moons may be detectable through perturbations of the position of the photocenter attributed to the ``planet" (method 5) and by measuring the doppler shift of the planet's spectrum due to motion about the planet-moon barycenter (method 6).  Assuming the planet's orbital orientation is known, through e.g. radial velocity and astrometry measurements, through methods 4-6, the moon's orbital orientation and relative orbital direction can, in principle, be determined.  In particular, for the case where the moon's orbit is face on, measurements of the photocenter allow the orbit direction to be determined, while for the case where the moon's orbit is inclined with respect to the plane of the sky, and motion along the line-of-sight is measureable, or mutual events are present and detectable, the moon orbital orientation can again be determined.  Such multi-epoch spectral and photometric monitoring follow-up may be conducted to investigate planetary properties, such as atmospheric composition, cloud structure and presence of rings \citep[e.g.][]{Green2003,Dyudina2005,Sudarsky2005} and a moon could be detected, along with its orbit direction, as a significant secondary return.  However, these methods were developed for telescopes with the capablilities of OWL \citep{Hook2005}, so again, the next generation of telescopes would be required.

\section{Summary and Discussion}

As discussed in section~\ref{Moon_Formation}, moon orbit direction is an important probe of moon formation pathway.  To take advantage of this information, two possible approaches can be employed.

One possible way to detect if a moon's orbit is prograde or retrograde is to use a moon detection method that is capable of determining this.  These are summarised along with the technology required to implement them for a range of planet-moon systems in table~\ref{SigTable}.  As a result, to use these techniques, to \emph{observationally} determine whether a moon's orbit is prograde or retrograde, the next generation of large telescopes is required.  

\begin{deluxetable}{lcccc}
\tabletypesize{\scriptsize}
\tablecaption{Signal sizes for the set of methods capable of determining moon orbit direction for a range of representative planet-moon systems.\vspace{-0.2cm}}

\tablewidth{0pt}
\tablehead{
\colhead{} & \colhead{Saturn-Titan} & \colhead{Jupiter-Saturn} & \colhead{Jupiter-Saturn} & \colhead{} \\ 
\colhead{Detection Method} & \colhead{($a_S$=$a_{Ti}$)} & \colhead{($a_S$=$a_{Ti}$)} & \colhead{($a_S$=$R_{H}$)} & \colhead{Ref.} 
}
\startdata
\textbf{Transiting Planets} &   &   &      &    \\
Direct measurement of  & 9.6$\times$10$^{-3} $kms$^{-1}$ \tablenotemark{b} & 2.1 kms$^{-1}$ \tablenotemark{b} & 0.72 kms$^{-1}$ \tablenotemark{b}     &   \citep[][sec. 2.1]{Cabrera2007} \\
planet line-of-sight velocity &  &  &  & (Average of planet and moon)\\
 \hline
 \textbf{RV Planets} &   &   &      &    \\
Perturbation of star & 3.3$\times$10$^{-7}$cms$^{-1}$ \tablenotemark{e} & 5.1$\times$10$^{-4}$cms$^{-1}$ \tablenotemark{e} & 0.87 cms$^{-1}$ \tablenotemark{e}     &   \citep[][eq. 23]{Morais2008}  \\
line-of-sight velocity due   &  &  &  & (Maximum value) \\
to planet-moon binarity &  &  &  & \\
\hline
  \textbf{Pulsar Planets} &   &   &      &    \\
Perturbation of pulsar & 2.4$\times$10$^{-6}$$\mu$s & 1.8$\times$10$^{-3}$$\mu$s & 74 $\mu$s \tablenotemark{a*}     &   \citep[][eq. 14]{Lewis2008} \\
time-of-arrival signal due   &  &  &  & \\
to planet-moon binarity &  &  &  & \\
\hline
  \textbf{Direct Imaging} &   &   &      &    \\
Motion of planet  & 1.4$\times$10$^{-2}$mas \tablenotemark{e} & 1.5 mas \tablenotemark{d} & 12 mas \tablenotemark{d}     &   \citep[][eq. 1]{Cabrera2007} \\
photocenter &  &  &  & \\
Radial velocity of  & 9.6$\times$10$^{-3}$kms$^{-1}$ \tablenotemark{e} & 2.1 kms$^{-1}$ \tablenotemark{d} & 0.72 kms$^{-1}$ \tablenotemark{d}     &   \citep[][sec. 2.1]{Cabrera2007} \\
planet &  &  &  & (Average of planet and moon)\\
Mutual events & 2.9$\times$10$^{-9}$ \tablenotemark{c} & 2.9$\times$10$^{-9}$ \tablenotemark{c} & 1.1$\times$10$^{-9}$ \tablenotemark{c}     &   \citep[][sec. 2.2]{Cabrera2007} \\
(rel. phot. precision) &  &  &  & \\
\enddata

\vspace{-0.6cm}

\tablecomments{The hosts star is a solar mass star at 10pc distance, and the planet-moon pair orbit 1AU from their host.  The moon's semi-major axis is that of Titan ($a_{Ti}$) or the planetary Hill radius ($R_H$).  Only one system is detectable with current technology (*), however, pulsar planets are rare.}
\tablenotetext{a}{Detectable with SKA \citep{Liu2011} \vspace{-0.15cm}}
\tablenotetext{b}{Not detectable with a 8.2m diameter telescope \citep{Snellen2010}\vspace{-0.15cm}}
\tablenotetext{c}{Detectable with a 30m diameter telescope \citep{Hook2005} \vspace{-0.15cm}}
\tablenotetext{d}{Detectable with a 100m diameter OWL-like telescope \citep{Hook2005} \vspace{-0.15cm}}
\tablenotetext{e}{Not detectable with a 100m diameter OWL-like telescope \citep{Hook2005,Lovis2006}\vspace{-0.15cm}}
\label{SigTable}
\end{deluxetable}

Alternatively, dynamics can be used to determine if a moon is in a prograde or retrograde orbit for some cases. If the satellite orbital elements can be measured to sufficient accuracy, and a sufficiently long time series is available, satellite orbital evolution due to three-body perturbation (distant moons) may differentiate between the cases.  However, all causes of orbital perturbation e.g. planetary oblateness \citep[e.g.][p. 264]{Murray1999}, must be included.  In addition, three-body stability may be used to determine orbit direction, as, moons in distant retrograde orbits can be three body stable while the equivalent prograde orbit would be unstable \citep[e.g][]{Hamilton1991}, so, any moon found in such a configuration must have a retrograde orbit.  For the case of transiting planets, the masses of the planet and moon, the orbital eccentricity, semi-major axis and projected orientation can be constrained from the transit ligthcurves \citep[e.g][]{Kipping2009a}, given sufficient data quality and number of transits, so dynamical constraints could be used to differentiate between prograde and retrograde orbits for certain systems.  However, for the case of microlensed planets, where only one event is observed, less information is available, so using this method is more difficult.  Also, as moons are predicted to form/capture into orbits close to their host planet \citep[e.g.][]{Sasaki2010,Ochiai2014}, these cases are likely to be rare.

So, for distant moon orbits, it is vital to consider orbital evolution, and the shape of the three body stability boundary \citep[e.g.][]{Mardling2008}, while for the case where dynamics cannot differentiate between prograde and retrograde moon orbits, the next generation of telescopes is required.

\acknowledgments

We would like to thank S. Ida, B. Sato, M. Kuzuhara, M. Nagasawa, A. Higuchi, D. Kipping and an anonymous referee for helpful comments which improved the quality of this paper.  This work was supported by JSPS KAKENHI Grant Number: 24-02764.

\end{document}